\documentclass[runningheads]{llncs}

\usepackage{booktabs} 
\usepackage{multirow}
\usepackage[T1]{fontenc}
\usepackage{graphicx}
\usepackage{subfigure}
\usepackage{algorithm}  
\usepackage{algorithmic}
\usepackage[title]{appendix}
\usepackage{enumitem}
\usepackage{hyperref}
\usepackage{array}
\usepackage{units}

\begin{document}
%
\title{Enhancing SCF with Privacy-Preserving and Splitting-Enabled E-Bills on Blockchain}
%
%
\author{Hao Yang\inst{1} \and
Jie Fu\inst{1,2} \and
Zhili Chen\inst{1} \and
Haifeng Qian\inst{1} }
\authorrunning{H. Yang et al.}
\institute{East China Normal University, Shanghai, China \\
\email{51215902172@stu.ecnu.edu.cn} \and
The Hong Kong Polytechnic University, Kowloon, Hong Kong Special Administrative Region \\
\email{71215902083@stu.ecnu.edu.cn}
}

\maketitle   

\begin{abstract}
Electronic Bill (E-Bill) is a rucial negotiable instrument in the form of data messages, relying on the Electronic Bill System (EB System). Blockchain technology offers inherent data sharing capabilities, so it is increasingly being adopted by small and medium-sized enterprises (SMEs) in the supply chain to build EB systems. However, the blockchain-based E-Bill still face significant challenges: the E-Bill is difficult to split, like non-fungible tokens (NFTs), and sensitive information such as amounts always be exposed on the blockchain. Therefore, to address these issues, we propose a novel data structure called Reverse-HashTree for Re-storing transactions in blockchain. In addition, we employ a variant of the Paillier public-key cryptosystem to ensure transaction validity without decryption, thus preserving privacy. Building upon these innovations, we designed BillChain, an EB system that enhances supply chain finance by providing privacy-preserving and splitting-enabled E-Bills on the blockchain. This work offers a comprehensive and innovative solution to the challenges faced by E-Bills applied in blockchain in the context of supply chain finance.

\end{abstract}
\section{Introduction}
Supply chain finance (SCF) emphasizes the importance of the industrial chain. In order to facilitate fund flow and lower financing barriers for small and medium-sized enterprises (SMEs) within SCF, the primary organization usually issues bills to SMEs based on off-chain transactions\cite{b5}. Typically, a large enterprise assumes the role of the primary organization in the SCF. However, traditional bills face limitations due to their physical nature\cite{b4}, making them challenging to store, cumbersome to circulate, and susceptible to alterations.  
The blockchain-based E-Bill is a novel approach to addressing these issues. E-Bill is a digital version of a traditional bill. Blockchain, a distributed ledger technology, enables easier storage, traceability, and immutability of E-Bill. Originating from the Bitcoin system in 2008\cite{b3}, blockchains can be classified into public, private, and consortium chains based on different access mechanisms\cite{b6}. Due to access permissions, consortium chains possess higher efficiency and security compared to public chains\cite{b25}, making them suitable for applications such as IoT\cite{b7}, data sharing\cite{b8,b21}, electronic voting\cite{b12,b24}, federal learning\cite{b20} and more. As a promising technology for managing SCF, E-Bills are stored and circulated on blockchain to ensure their security.

However, there are always some challenges in integrating SCF and blockchain\cite{b6}. On the one hand, privacy security should be considered when applying blockchain to E-Bills. Due to the transparency of the data on the chain, miners can easily validate the legality of E-Bills and transactions, but sensitive information is also publicly released across all nodes. Sensitive information, such as amounts, could be exploited by attackers to compromise users' interests or assets. However, we cannot directly encrypt data within the blockchain, because we also want other parties to verify the legality of transactions in public scenarios. On the other hand, E-Bills resemble non-fungible tokens (NFTs), both possessing unique properties and being challenging to split\cite{b4}. In reality, the amount of an E-Bill may not exactly match the required transaction when an SME needs to make a payment. Thus, E-Bills must be splittable, but no known E-Bill splitting methods existed previously.


In this paper, we propose BillChain, a novel blockchain-based EB system that protects the amounts of E-Bills and enables flexible E-Bill splitting. The amounts of E-Bills are encrypted using homomorphic encryption to hide them. Homomorphic encryption is an encryption technique where specific transformations on ciphertext can be mapped to transformations on the plaintext. Due to a single structure and excellent execution efficiency, we choose the Paillier cryptosystem\cite{b9} as our fundamental encryption scheme. The challenging problem of composite residual classes forms the basis for the Paillier cryptosystem, which satisfies addition homomorphism. Other popular homomorphic encryption schemes at the moment include RSA partial homomorphism, ElGamal partial homomorphism, and the Gentry ideal lattice fully homomorphic encryption scheme. However, RSA and ElGamal are both multiplicative homomorphisms, while our scheme requires an additional homomorphism to guarantee the equality of transaction inputs and outputs. As a result, we use a variant of Paillier that allows publicly verifying the equality of transaction inputs and outputs in ciphertext. In addition, to address the problem of E-Bill splitting, we designed a data structrue called Reverse-HashTree (RHT) to store transactions in this paper. We achieve E-Bill splitting through RHT, which can also be applied to split other NFTs.

Finally, we implement BillChain on Hyperledger Fabric. Specifically, we run more than 50,000 transactions to evaluate the practicality and feasibility by throughput per second and transaction latency. Additional experiments are performed to find the best configuration parameters of Fabric, such as MaxMessageCount and BatchTimeout. The experimental results show that at up to 50,000 transactions, the optimal TPS of BillChain remains above 1,000, and the latency is typically kept under 2 seconds, which demonstrates that BillChain can efficiently process massive transactions.

In summary, we have made the following contributions:
\begin{enumerate}
\item[$\bullet$] We proposed BillChain, a novel blockchain-based E-Bill system that protects the amount privacy of E-Bills and allows for flexible splitting of E-Bills by storing them in RHT.
\item[$\bullet$] We designed a data structure called a Reverse-HashTree, which can be used to solve the problem of E-Bill splitting. RHT can also naturally resist and detect double-spending attacks. In addition, we proposed a special splitting scheme to guarantee the legality of splitting E-Bills and analyzed the security of the scheme.
\item[$\bullet$] We evaluated the system, and the results demonstrate its practicality and feasibility.
\end{enumerate}

The rest of this paper is structured as follows: Section 2 introduces related work. Section 3 provides the background. Section 4 provides an overview of BillChain. Section 5 details the splitting scheme. We discuss security in Section 6 and the implementation of the system in Section 7. In Section 8, we conclude this work.

\section{Related Work}
To protect users' privacy on the blockchain, researchers have proposed some methods, including encryption to hide sensitive information and mixing services to confuse the address link relationship. In 2013, MaxWell\cite{b18} proposed the first decentralized mixing scheme, Coinjoin, but the scheme can only confuse the address link relationship, guaranteeing the anonymity of the identity but not obscuring the transaction amounts. Ni et al.\cite{b10} then conducted a series of work on mixing services, lowering the cost of the service but leaving transaction amounts exposed. Zerocash\cite{b12} introduced non-interactive zero-knowledge proof to protect both parties' addresses and transaction amounts, but it is inefficient because it relies entirely on non-interactive zero-knowledge (NIZK) proof. Q.Tang\cite{b22} designed Public key encryption schemes supporting equality test with authorisation of different granularity. H. Lipmaa\cite{b23} proposed verifiable homomorphic oblivious transfer and private equality test, supporting to verify equality without threatening privacy. DumbAccount\cite{b14} scheme hides the transaction amounts using homomorphic encryption and zero-knowledge proof, but it cannot meet both parties' needs for identity privacy. It is worth pointing out that no known E-Bill splitting methods existed previously. Our solution offers significant advantages over other related work in terms of comprehensive privacy protection, system efficiency, and meeting practical application needs, achieving a balance between privacy and practicality.

\section{Preliminaries}
\subsection{Fabric and Chaincode} 
Hyperledger Fabric\cite{b2} is an open-source, enterprise-grade distributed ledger platform created by the Linux Foundation. It was originally designed for enterprise-level applications. Unlike popular public chain platforms, Hyperledger Fabric is a consortium chain for enterprise applications. It has many different features, including permission control, private data, and chaincode. For example, private data provides privacy preservation, allowing enterprises to share only the data they wish to share among authorized participants. In addition, chaincode automates business processes based on self-enforcing clauses that companies write into chaincode. The code and protocols exist in a decentralized blockchain network, transactions are traceable and irreversible, and Fabric creates trust between organizations. This allows enterprises to make smarter decisions, save time, and reduce costs and risks.

\subsection{Cryptographic Primitives}
The cryptographic building blocks consist of the following components: Paillier encryption scheme, non-interactive zero-knowledge proofs, and digital signature schemes. We informally describe these notions below:


\noindent \textbf{Paillier Encryption Scheme.} The basic paillier is thoroughly described in the literature\cite{b17}, and it is made up of three algorithm groups $(KGen, Enc, Dec)$ , which represent key generation, encryption, and decryption algorithms, respectively.  The $KGen$ is used to generate public and private key pair $(pk,sk)$.  For a given message $m$, corresponding ciphertext $c$ can be obtained by $Pk$ and encryption algorithm $E(m)=g^m r^n \, mod \, n^2 $.  Pailiier has the additive homomorphic property:

\begin{equation}
Dec_{sk}(Enc_{pk}(m_1) \cdot Enc_{pk}(m_2) \bmod\, n^2)=m_1+ m_2 \, \bmod \, n\label{eq}
\end{equation}

However, due to encrypting the output and input using various random numbers, the basic paillier is unable to satisfy the requirement of verifying that the transaction's output and input are equal under ciphertext.  For example, given an input $in$ and $n$ outputs $out_i (i=1,2...n)$ that satisfy $in = \sum_{i=1}^n \limits out_i$. Then corresponding ciphertexts could be calculated using randomly numbers $r_{in},r_i \in Z_{n^2}^{*}$:

\begin{equation}
    C_{in}=g^{in} r_{in}^n \, \bmod \, n^2\label{eq}
\end{equation}

\begin{equation}
    C_{out_i}=g^{out_i} r_{i}^n \, \bmod \, n^2\label{eq}
\end{equation}

Next we can calculated 
\begin{equation}
   c_{in'} = \prod_{i=1}^{n} c_{out_i} = g^{\sum_{i=1}^{n} out_i} \cdot \left(\prod_{i=1}^{n} r_i\right)^{n} \bmod n^2 \label{eq}
\end{equation}

But $c_{in'}$ doesn't necessary equal to $\prod_{i=1}^{n} c_{out_i}$ as $r_{in} \neq \prod_{i=1}^{n} \limits r_i$, we couldn't verify the equality relation between the input and output in a transaction under the ciphertext.  Therefore, we use a variant of paillier as following: 

\begin{itemize}
    \item[$\bullet$] $KeyGen(t,l,s,T) \Rightarrow (pk_i, pk_w)$ \\
        The input is security parameters for encryption and verification separately. The outputs are the public key of every nodes i with the private key $sk_i$ and the public key $pk=(n,g)$.  $pk_w$ is whole network's homomorphic public key with the private key $sk_w$ which is only held by the primary organization.  	\\
    \item[$\bullet$] $Encrypt(in, out_i) \Rightarrow (c_{in}, c_{out_i})$	\\
        The inputs are plaintext of transaction's a input $in$ and multi outputs $out_i$, then randomly select numbers $r_{in},r_i$  satisfy $r_{in} = \prod_{i=1}^n \limits r_i$. The outputs are $c_{in}=g^{in} (r_{in})^{n} \bmod n^2,c_{out_i}=g^{out_{i}} (r_{i})^{n} \bmod n^2$ \\
    \item[$\bullet$] $Dec \Rightarrow (c_{in}, c_{out_i})$
    \item[$\bullet$] $Verify(c_{in}, c_{out_i}) \Rightarrow 1/0$	\\
        This algorithm calculates 
        \begin{equation}
               c_{in'} = \prod_{i=1}^{n} c_{out_i} = g^{\sum_{i=1}^{n} \limits out_i} \cdot \left(\prod_{i=1}^{n} r_i\right)^{n} \bmod n^2 \label{eq}
        \end{equation}
        then outputs 1 if $c_{in} = c_{in}'  $or 0 otherwise.
\end{itemize}

As mentioned above, we limit the selection of random numbers, allowing the equality relation between the input and output in a transaction under ciphertext. This enables us to verify transaction amounts without any trusted third party and without revealing the plaintext.

\noindent \textbf{Interval proof.} The idea of determining whether a committed integer falls within a given range was first proposed in\cite{b19} and further expanded in\cite{b13,b15}. The CFT Proof\cite{b16} achieves the goal of demonstrating that a secret value is within a given range. However, the CFT proof has a very high expansion rate. The protocol is too complex to be used in our scheme, but it can limit the committed number to a range [a, b] with minor deviation. Our goal is to demonstrate that the encrypted quantities are between [0, m], so we use Wu's approach\cite{b15}, which is both effective and practical.

\subsection{Reverse-HashTree.}
The Merkle tree is known to be used to store transaction data and account The Merkle tree is widely used in most blockchains to store transaction data and account status. The hash value is obtained by hashing transactions separately and then combining two selected hash values into a new hash value, repeating this process. Finally, the Merkle tree root is obtained and stored in the block header, while the Merkle tree is kept in the block body. However, the bottom-up design of the Merkle tree is not suitable for splitting E-Bills.

\begin{figure}
\includegraphics[width=\textwidth]{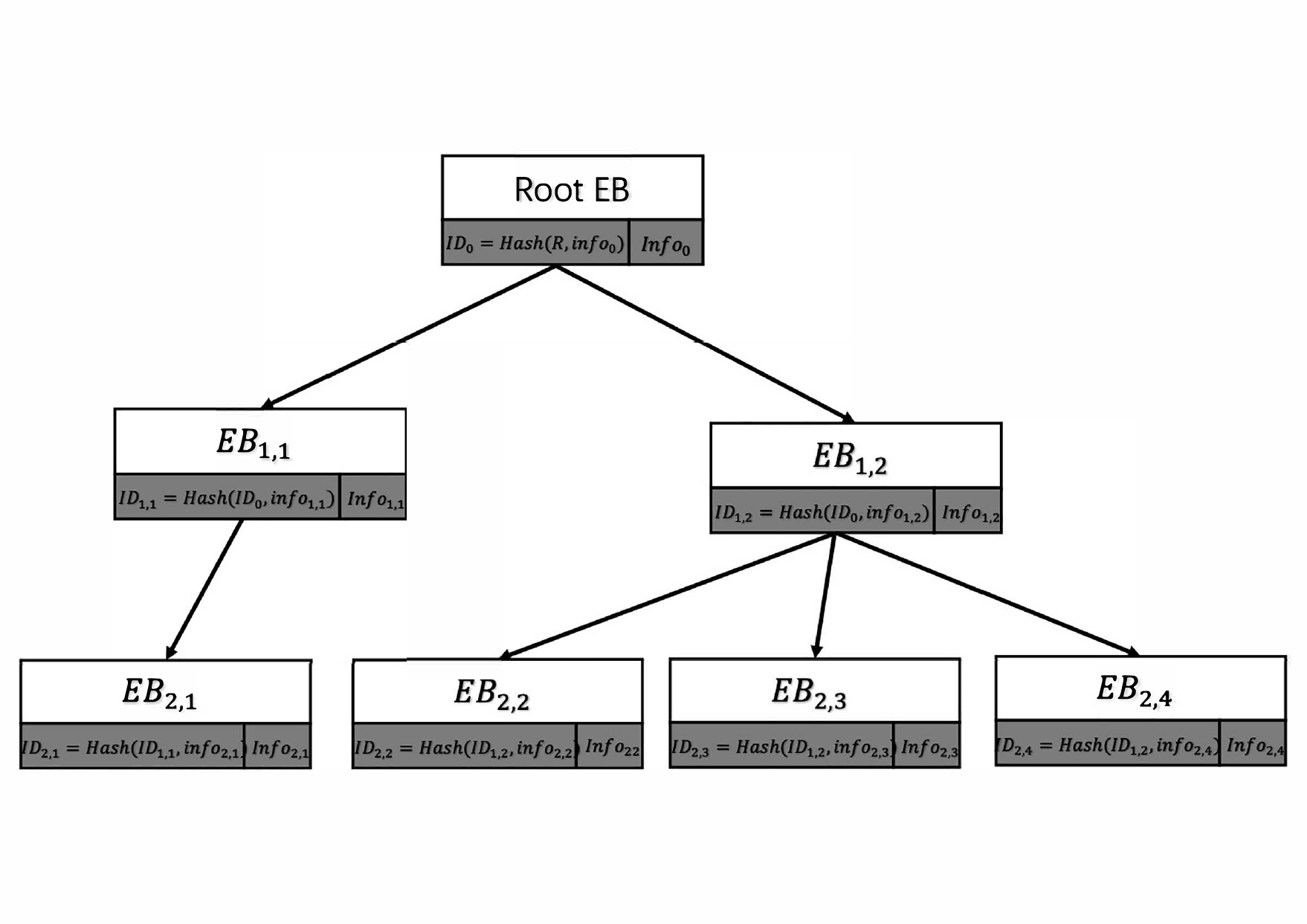}
\caption{Reverse HashTree.} \label{fig1}
\end{figure}

To flexibly split E-Bill, we propose using the Reverse-HashTree (RHT) to store E-Bill. Figure \ref{fig1} displays the RHT structure, which is a tree-like structure linked by hash functions. RHT ensures the security of E-Bill splitting through the one-way linkage of hash functions. Each RHT node data comprises two parts: the ID and the INFO of the E-Bill. ID is a unique 256-bit number and can be obtained by calculating: $id=Hash(id_{pre}, info)$, where $id_{pre}$ is the ID of the parent node and $info$ is the information of the E-Bill. Specifically, the Root E-Bill selects a random number $R$ to replace $id_{pre}$ to obtain the ID. $info$ contains multiple attribute fields: $$info= \{amount, owner, path\}$$ where \textit{amount} represents the value of the E-Bill, \textit{owner} is the current owner of the E-Bill, and \textit{path} is a hash path that represents the position of the E-Bill in the RHT, denoted by a set of IDs. For example, the \textit{path} of $EB_{3,1}$ is $\{id_0, id_1, id_3\}$.

In BillChain, the primary organization issues E-Bill for SME organizations based on off-chain transactions, and E-Bills are subsequently transferred or split, generating new E-Bills that circulate on the blockchain. Since splitting transactions may have multiple outputs to generate E-Bills, RHT is not always a binary tree. The root node of the RHT stores the initially issued E-Bill, known as the Root E-Bill. When the E-Bills are transferred or split, child nodes extend downward from the root node to store the generated E-Bills, known as Sub-Bills.

Based on the constructing process, RHT has the following features:
\begin{enumerate}
    \item[$\bullet$] All currently unspended E-Bill are stored in RHT leaf nodes;
    \item[$\bullet$] E-Bill based on RHT  are inherently immune to double-spending attacks;
    \item[$\bullet$] An RHT includes the entire life cycle of a Root-Bill circulation from issue to end.
\end{enumerate}

\section{A model proposal:an overview}
\begin{figure}[htbp]
\centerline{\includegraphics[width=0.92\textwidth,height=0.72\textwidth]{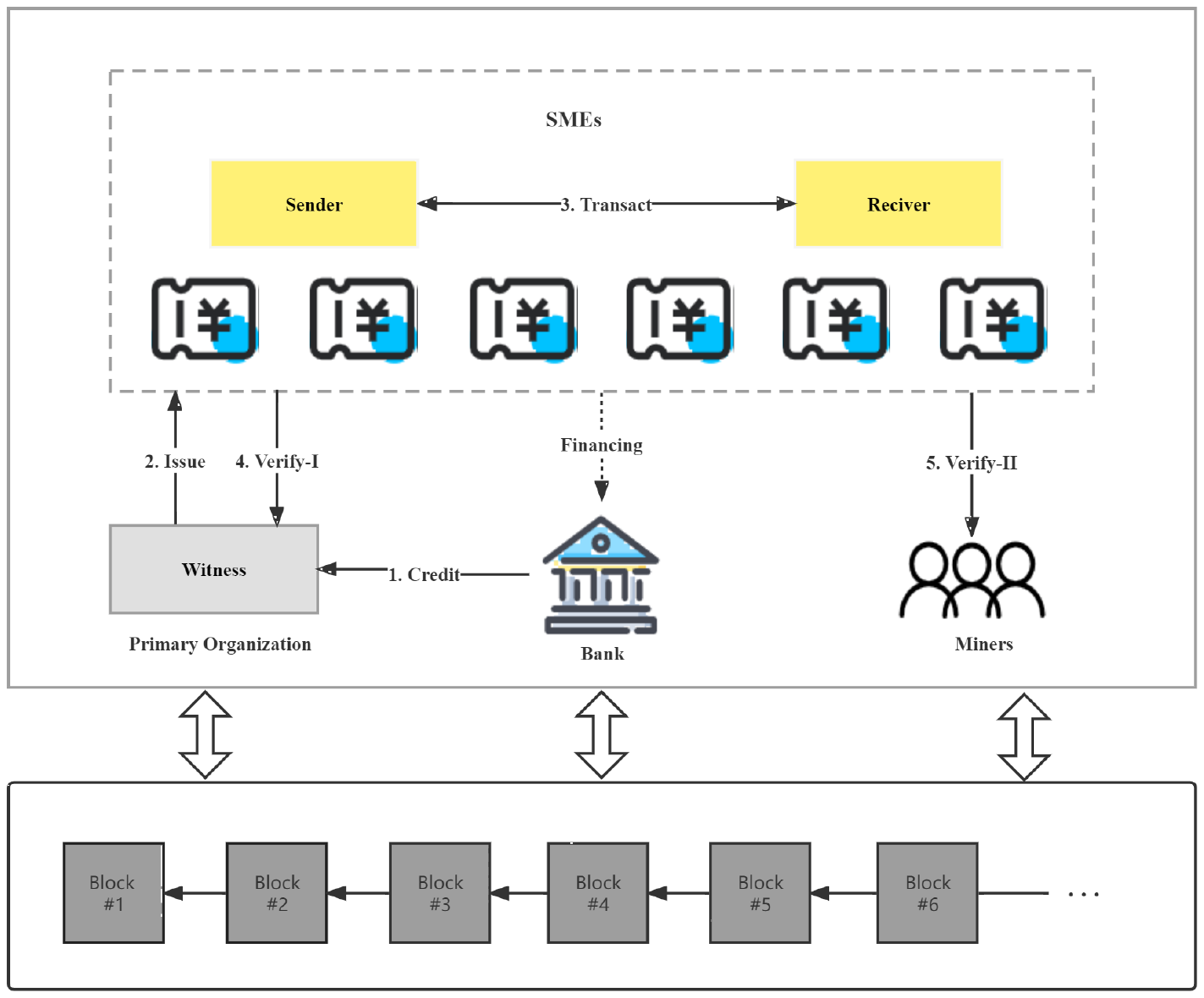}}
\caption{Blockchain-based E-Bill system model}
\label{fig2}
\end{figure}

BillChain employs a consortium chain as the system's framework and facilitates the circulation of E-Bills in BillChain. As illustrated in Figure \ref{fig2}, the bank, acting as a regulatory organization, provides credit for the primary organization. Subsequently, the primary organization issues E-Bills to SME organizations under the bank's supervision, and multi-level SME organizations receive, transfer, and split the E-Bills. SME organizations can also request financing from banks using the E-Bills they own, which lowers the financing threshold for SME organizations. The following section defines the roles and Corresponding functions in a transation.

\noindent \textbf{Sender:} The sender $S$ of a transaction can only be one, and it can be any node managed by the SME organization or primary organization. $S$ is responsible for initiating a transaction request.

\noindent \textbf{Receiver:} The receiver $R$ of a transaction could be multiple and can be any number of nodes managed by the SME organization in BillChain. The transfer transaction has only one receiver, while the splitting transaction usually has two or more. $R$ is responsible for checking whether the transaction amount equals the amount the receiver wants to receive and then sign the transaction if they decide to accept it.

\noindent \textbf{Witness:} The witness $W$ is the only E-Bill issuer and the holder of the entire network's homomorphic public and private keys, corresponding to the primary organization in SCF. $W$ is responsible for auxiliary checking of transaction amounts and issuing E-Bills. It is worth noting that we do not view the witness node as a completely trustworthy party but rather only check whether the transaction amount equals the amount the receiver should receive under ciphertext. Therefore, $W$ does not have the ability to obtain the transaction amount and tamper with the transaction.

\section{The Proposed Scheme}
\subsection{System Setup}
A set of users $\mathcal{U} = {U_1, U_2, ...}$, an authenticated witness $\mathcal{W}$, and a set of miners $\mathcal{M}$. With the security parameter $pp$, the system parameters ${p, G, E, H}$ are determined, where $G$ is an additive group with generator $p$, $E$ is a secure symmetric encryption algorithm, and $H$ is a secure hash function $H: {{0, 1}}^* \to G$. $\mathcal{W}$ randomly chooses $s \in Z_p^*$ and computes $pk_w = sP$, $k = h(s)$. The secret keys of $\mathcal{W}$ are $sk_w = (s, k)$, and the corresponding public key is $pk_w$. Every user $i$ has their own public and private key pair $(pk_i, sk_i)$ generated by $KeyGen()$.


\subsection{Transaction Workflow} 
\begin{figure}[htbp]
\centerline{\includegraphics[width=0.92\textwidth]{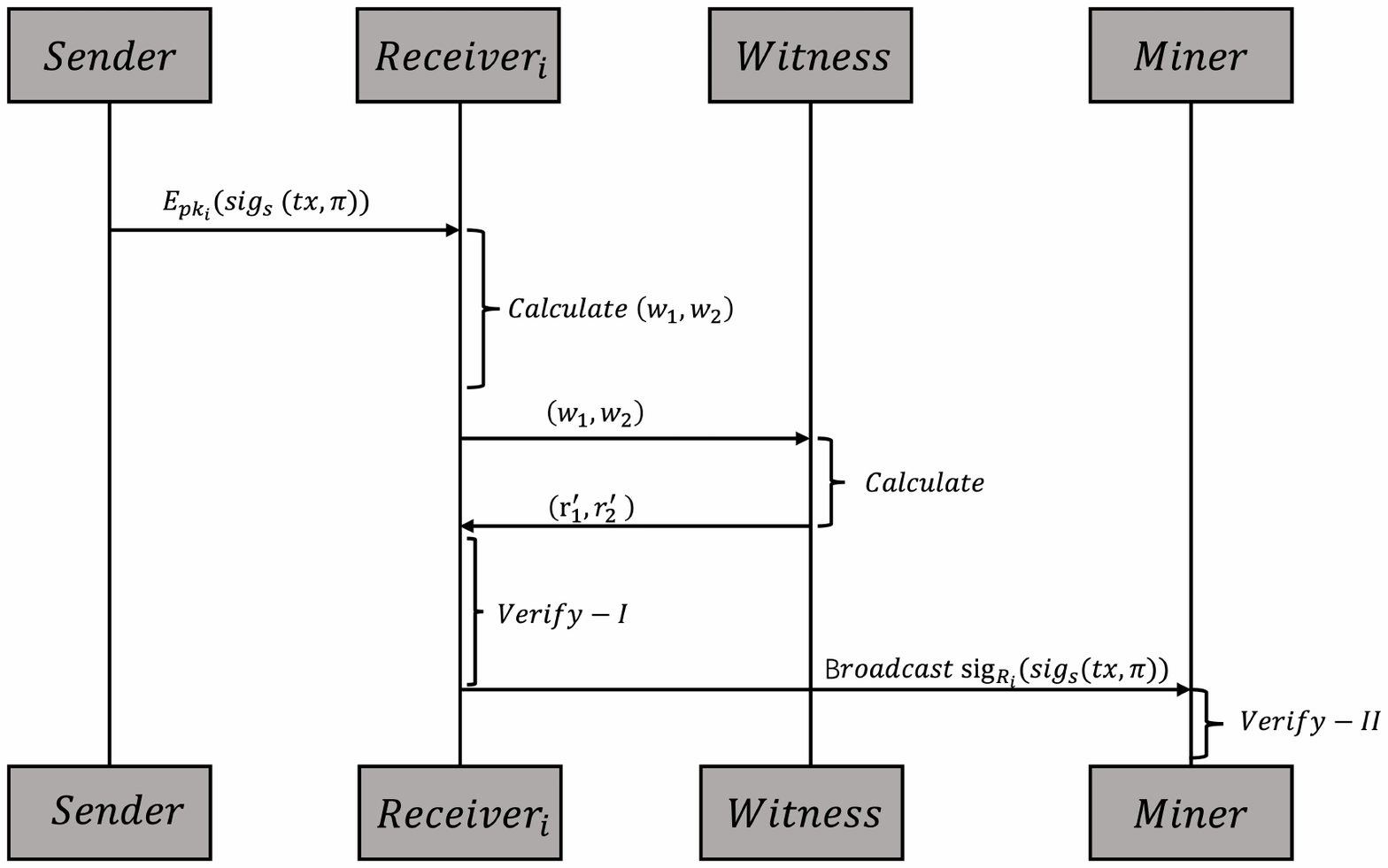}}
\caption{Transaction workflow}
\label{fig3}
\end{figure}

Figure \ref{fig3} illustrates the complete transaction workflow. The sender $U_S$ ($U_S \subseteq \mathcal{U}$) initiates a transaction $tx$ with a proof $\pi$, which is an interval proof to ensure that all outputs and inputs in $tx$ are greater than 0. Then, $U_S$ creates a digital signature using its private key $sk_S$ and encrypts $sig_S(tx,\pi)$ to send it to each receiver $U_{R_i}$. Upon receiving the transaction, $U_{R_i}$ can verify the transaction amount with the assistance of witness $\mathcal{W}$. For instance, if $U_{R_i}$ is supposed to receive the amount $out_i$, but the amount actually received in $tx$ is $out_i'$, it needs to confirm that $out_i$ and $out_i'$ are equal. If the verification passes, $U_{R_i}$ uses its private key $sk_{R_i}$ to sign and obtain $sig_{R_i}(sig_S(tx,\pi))$. Finally, after receiving $sig_{R_i}(sig_S(tx,\pi))$, miner nodes further ensure the transaction's validity by performing the subsequent five-step verification. If all verifications pass, nodes in the network select a final block and add it to the blockchain according to a consensus mechanism. The transaction process does not reveal any information about the transaction amount.



\subsubsection{Initial Transaction}     
\paragraph{\textbf{Sender:}} $U_S$ initials a transaction for splitting E-Bill. \\
When $U_S$ initials a transaction to split a E-Bill $\mathcal{E}$ to ${U_R}_i$, i.e., $U_1$ splits the E-Bill $\mathcal{E}$ to $U_2,U_3$, the input amount is credited as $in$,the split amounts are credited as $out_1$ and $out_2$.   \\
-  Sender $U_S$ proposes a splitting transaction $tx$:   

\begin{equation}
    tx = \{{E_{pk_w}}(in), {pk_i:E_{pk_w}(out_i)}, \mathcal{E}\}
\end{equation}

where $in$ is the amount of E-Bill $\mathcal{E}$ to be split and $pk_i$ are receivers' public key. $\mathcal{E}$ is the E-Bill to be split, and the corresponding owner can be found according to its path information in RHT as above referring. All input and outputs are encrypted by $E$ using whole network homomorphic public key $pk_w$ \\
- $U_S$ constructs an interval proof $\pi$ for proving each output $out_i>0$ using the scheme\cite{b15}.  \\
- $U_S$ signs the transaction $m$ using his private key $sk_s$. \\
- $U_S$ encrypts $m$ separately using $U_{R_i}$'s publick key $pk_{i}$ to get  $E_{pk_i}(sig_S(tx,\pi))$.     \\
- $E_{pk_i}(sig_S(tx,\pi))$ is sent to each receiver ${U_R}_i$.        \\

When ${U_R}_i$ received the ciphertext $E_{pk_i}(sig_S(tx,\pi))$, he firstly decrypts it  to obtain $sig_S(tx, \pi)$  then verify the signature using sender's public key and the sender's public key can be found in $\mathcal{E}$.  If verification pass, ${U_R}_i$  need further to confirm by witness if the amount $out_i$ he received is equal to the amount $out_i^{'}$ he should receive.

\subsubsection{Amount checking(Verify-I)}  
\paragraph{\textbf{Receiver}}: ${U_R}_i$ verify transaction amount by Witness $\mathcal{W}$ \\
When ${U_R}_i$ get $sig_S(tx,\pi)$, he can't know the specific amount he received in fact, because the $out_i$ was encrypted by $\mathcal{W}$'s public key $pk_w$. So he need to confirm the real amount he received. The process is as follows: \\
- ${U_R}_i$ randomly chooses $r_1,r_2$ to verify if $out_i$ is equal with $out_i^{'}$, where $out_i$ is the amount in $tx$ and $out_i^{'}$  is the amount which he should receive. \\
- ${U_R}_i$ calculates $(w_1, w_2)$ as below:

\begin{equation}
  w_1=out_i \div out_i^{'} \times E_{pk_w}(r_1)\label{eq}
\end{equation}
\begin{equation}
  w_2=E_{pk_w}((r_2)\label{eq}
\end{equation}

- ${U_R}_i$ sends $(w_1, w_2)$ to $\mathcal{W}$, and $\mathcal{W}$ decrypts it using the whole network homomorphic private key $sk_w$:

\begin{equation}
  r_1'=Dec_{pk_w}(w_1)\label{eq}
\end{equation}
\begin{equation}
  r_2'=Dec_{pk_w}(w_2)\label{eq}
\end{equation}

- $\mathcal{W}$ obtains $(r_1',r_2')$ and returns them to ${U_R}_i$. ${U_R}_i$ continue to verify as below:

\begin{equation}
  Equal\{(r_1,r_2),(r_1',r_2') \}=1/0\label{eq}
\end{equation}

The output is 1 if $r_1=r_1'$ and $r_2=r_2'$, or 0 otherwise. $R_i$ accepts the transaction when the output is 1, then uses its private key $sk_{R_i}$ to sign and obtain $sig_{R_i}(sig_S(tx,\pi))$, and subsequently broadcasts it in the blockchain network. Otherwise, ${U_R}_i$ rejects the transaction $tx$.

\subsubsection{Miner verification(Verify-II)} 
\paragraph{\textbf{Miners: }}$\mathcal{M}$ publicly verifies the transaction validity and reaches consensus to package the transaction into a block. \
Upon receiving $sig_{R_i}(sig_S(tx,\pi))$, miner nodes further ensure the transaction's validity by performing the following five steps:

\noindent \textbf{Verify-1: }Verify the sender and receiver's signatures. The miner node uses the public keys $pk_{R_i}$ and $pk_S$ to verify the signatures, determining whether the digital signatures are provided by the sender and receivers. If Verify-1 passes, proceed to Verify-2.

\noindent \textbf{Verify-2: }Verify double spending and E-Bill ownership. Using the 'path' field in the $msg$, determine the position of the E-Bill $\mathcal{E}$ in the RHT, then verify whether the position is a leaf node of RHT, and confirm if the owner of $\mathcal{E}$ is the sender. Based on the properties of RHT, if the node is not a leaf node, it proves that the E-Bill $\mathcal{E}$ has been spent. If Verify-2 passes, proceed to Verify-3.

\noindent \textbf{Verify-3: }Verify that the E-Bill has not been tampered with. The 'Path' in $tx$ must match the path in the RHT, as the path will change if the E-Bill is tampered with. If Verify-3 passes, proceed to Verify-4.

\noindent \textbf{Verify-4: }Verify the correctness of the transaction amounts. Check by the interval proof $\pi$ if each transaction output amount is greater than 0. If Verify-4 passes, proceed to Verify-5.

\noindent \textbf{Verify-5: }Verify the legality of the transaction amounts. The legality of the transaction amounts refers to the transaction input and output amounts being equal. It can be verified that the following equation holds true:

\begin{equation}
  E_{pk_w}(in)=\prod_{i=0}^nE_{pk_w}(out_i)\label{eq}
\end{equation}

If $m$ five steps are passed, the miner node will accept $m_i$, and multiple miner nodes will agree on the transaction and upload $m_i$ to the chain.  On the contrary, if a error occurs in any of the above five steps of verification, miner node will reject the transaction and send an error message to the sender and receivers.



\section{Security Analysis}
\noindent \textbf{Double spending attack.}
In RHT, only the E-Bills stored in the leaf nodes are considered valid, as they have not been spent when prior transactions are lawful. If the transaction sender initiates a transaction repeatedly, Verify-1 will not pass. \
\noindent \textbf{Collusion attack.}
In BillChain, there are three possible ways collusion attack might take place: between the sender and the witness, between the sender and the receiver, or between the witness and the receiver.
\begin{enumerate}
\item Sender and witness collude: If the sender forges the transaction amounts $out_i<0$, the receiver may approve the Verify-1 result, but the miner node cannot confirm the legitimacy of the output amounts in Verify-4, or the input and output of a transaction cannot be equal in Verify-5;
\item Sender and receiver collude: When the sender fakes a non-existent E-Bill to the receiver, the miner node cannot find the E-Bill to be traded through Path in Verify-2;
\item Witness and receiver collude: If the receiver tampers with the transaction amounts, the miner node cannot pass Verify-5 by computing the input and output of the transaction.
\end{enumerate}
BillChain does not intend to address the malicious behavior of miners, as it is restricted by the blockchain platform and consensus algorithm.\

\noindent \textbf{Amounts privacy.}
By encrypting the amounts of E-Bills and uploading them to the blockchain, BillChain achieves public verifiability of E-Bill amounts in ciphertext, based on a variant of the Paillier cryptosystem. The amounts cannot be decrypted without knowing $sk_w$. A single receiver also cannot know the received amounts of other receivers in the same transaction.

\section{Implementation}
In this section, we design experiments to evaluate the usability and feasibility of BillChain. BillChain runs on a system with 16-core 32GB memory, and Linux using Ubuntu20.04. Based on the widely used open-source consortium chain framework, Hyperledger Fabric, we build a blockchain network consisting of 2 peer nodes and 1 order node. Peer nodes in Fabric are responsible for simulating smart contract execution and endorsing requests from other nodes. The order nodes sort the read-write sets generated by smart contract execution according to the endorsement policy.

To determine the ideal blockchain configuration, we first test BillChain's performance under various blockchain network configurations. In the configuration process, the primary indicators that affect blockchain performance include MaxMessageCount and BatchTimeout\cite{b1}. MaxMessageCount specifies the maximum number of transactions included in each block, and BatchTimeout specifies the maximum block delay. When MaxMessageCount or BatchTimeout are reached, the current transactions are packaged to generate a new block. TPS (Transactions Per Second) and latency are common indicators of blockchain performance. TPS is the average number of transactions processed by the blockchain network each second, and latency is the time it takes for a transaction to be recorded from the moment it is submitted.

\begin{table}[htbp]
\renewcommand{\arraystretch}{1.25}
    \caption{The performance under various blockchain configurations}
    \label{tab1}
    \centering
        \resizebox{0.6\textwidth}{!}{
        \begin{tabular}{*{7}{c}}
            \toprule
            \multicolumn{1}{c}{\multirow{2}*{Type}} & 
            \multicolumn{1}{c}{\multirow{2}*{BatchTime(s)}} & 
            \multicolumn{5}{c} {MaxMessageCount} \\
            
            \cmidrule(lr){3-7}
            \multicolumn{2}{c}{~} & 1 & 8 & 16 & 32 &64 \\
            \midrule 
            \multirow{4}*{TPS} 
             & 0.2  & 49.5 & 80.6 & 82.5 & 83.1 & 82 \\
             & 1   & 51.4 & 80.8 & 84.7 & 83,7 & 78.8 \\
             & 2   & 48.9 & 80.0 & 83.4 & 83.8 & 82.2 \\
             & 3   & 48 & 79.3 & 83.3 & 83.9 & 82.5 \\
             \hline
             \multirow{4}*{Latency(s)} 
             &0.2 & 1.56 & 0.37 & 0.29 & 0.3 & 0.29 \\
             & 1 & 1.63 & 0.41 & 0.36 & 0.33 & 0.4 \\
             & 2 & 1.72 & 0.38 & 0.3 & 0.34 & 0.4 \\
             & 3 & 1.8 & 0.34 & 0.32 & 0.33 & 0.52 \\
            \bottomrule  
        \end{tabular}
        }
\end{table}

Table a and b in Table \ref{tab1} respectively demonstrate the changes in transactions per second (TPS) and Latency with various combinations of MaxMessage and BatchTimeout. The TPS shows that when MaxMessage=1, each transaction is individually packaged into a block, resulting in greater storage redundancy and lower TPS. As MaxMessage increases, the transaction volume contained in a block also increases, reducing the relative communication overhead and causing the TPS to gradually increase and stabilize. The latency indicates that when MaxMessage is small and BatchTimeout is large, the block generation rate is slow and latency is high. Taking into consideration the improvement of TPS, reduction of Latency, and minimizing communication overhead, we ultimately choose the blockchain network configuration with MaxMessageCount=16 and BatchTimeout=2. 

\begin{figure*}[htbp]
\centerline{\includegraphics[width=1.2\textwidth]{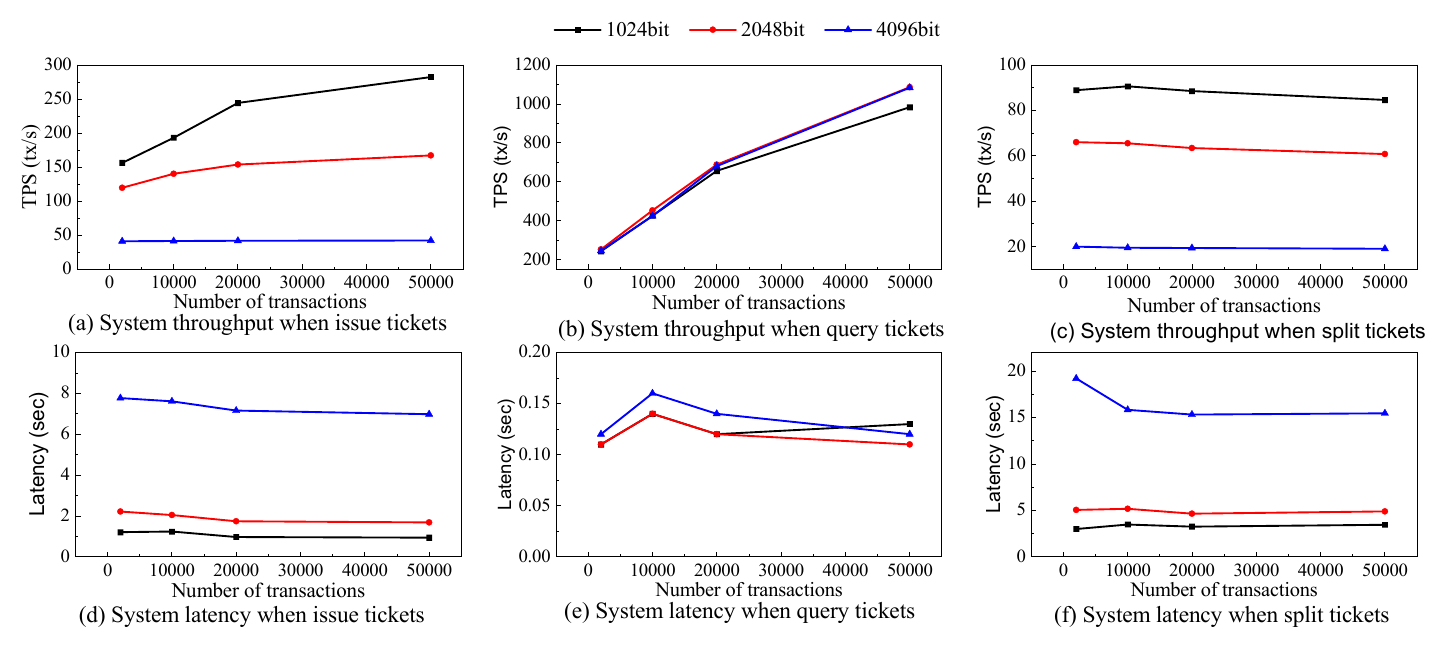}}
\caption{Results of varying key lengths and transactions}
\label{fig4}
\end{figure*}

Next, we evaluate the TPS and latency of BillChain with key lengths of 1024 bits, 2048 bits, and 4096 bits. We use the open-source blockchain testing tool Hyperledger Caliper\cite{b1} to test the performance. To simulate BillChain in the test environment, we randomly tested the performance with 2,000, 10,000, 20,000, and 50,000 transactions, respectively.

As shown in Figure \ref{fig4}(a-c), for the TPS in BillChain with different transaction numbers and different key lengths, the key length significantly impacts the TPS. As illustrated in Figure a, when the key length is 1024 bits and the number of transactions is 50,000, the highest TPS is 282 for the issue transactions. However, when the key length is 4096 bits and transactions are 50,000, the TPS of the issue transactions is only 42.3. Furthermore, we find that as key length and transactions increase, the TPS of the query and issue transactions gradually increase, while the TPS of split transactions remains stable due to the large CPU and IO overhead involved in split Bill transactions. As shown in Figure \ref{fig4}(d-f), the latency of the issue and query transactions in BillChain is generally less than 2 seconds, while the latency of split transactions is relatively high, reaching up to 19.2 seconds. The latency is acceptable in general because the split transactions are fewer than the issues and query transactions in actual scenarios. 

\section{Conclusion}
In this paper, we have proposed a blockchain-based EB system to protect the amounts of E-Bills and realize the splitting of E-Bills. Specifically, we have employed a variant of the Paillier cryptosystem to achieve public verifiability of E-Bill amounts in ciphertext while preserving the privacy of the amounts. Moreover, to tackle the challenge of splitting E-Bills, we have designed a novel data structure, the Reverse-HashTree (RHT). The RHT enables efficient splitting of E-Bills with arbitrary amounts and can be extended to splitting other non-fungible tokens (NFTs). By conducting experiments to evaluate performance, BillChain demonstrates that it meets the needs of the E-Bill service in SCF and strengthens transaction security.


%
%
%
%

\end{document}